\documentclass[aps, prb, reprint, superscriptaddress, preprintnumbers, amsmath, amssymb, letterpaper, floatfix]{revtex4-1}

\usepackage{graphicx}
\usepackage[first]{draftcopy}

\begin{document}

\preprint{}

\title{\boldmath Persistent spin dynamics and absence of spin freezing in the $H$-$T$ phase diagram of the 2D triangular antiferromagnet YbMgGaO$_4$}

\author{Zhaofeng Ding}
\author{Zihao Zhu}
\author{Jian Zhang}
\author{Cheng Tan}
\author{Yanxing Yang}
\affiliation{State Key Laboratory of Surface Physics and Department of Physics, Fudan University, Shanghai 200433, China}
\author{Douglas E. MacLaughlin}
\affiliation{Department of Physics and Astronomy, University of California, Riverside, California 92521, USA}
\author{Lei Shu}
\altaffiliation[Corresponding author: ]{leishu@fudan.edu.cn}
\affiliation{State Key Laboratory of Surface Physics and Department of Physics, Fudan University, Shanghai 200433, China}
\affiliation{Collaborative Innovation Center of Advanced Microstructures, Nanjing University, Nanjing 210093, China}
\affiliation{Shanghai Research Center for Quantum Sciences, Shanghai 201315, China}

\date{\today}

\begin{abstract}
We report results of muon spin relaxation and rotation ($\mu$SR) experiments on the spin-liquid candidate~YbMgGaO$_{4}$. No static magnetism $\gtrsim 0.003\mu_B$ per Yb ion, ordered or disordered, is observed down to 22~mK, a factor of two lower in temperature than previous measurements. Persistent (temperature-independent) spin dynamics are observed up to 0.20~K and at least 1~kOe, thus extending previous zero-field $\mu$SR results over a substantial region of the $H$-$T$ phase diagram. Knight shift measurements in a 10-kOe transverse field reveal two lines with nearly equal amplitudes. Inhomogeneous muon depolarization in a longitudinal field, previously characterized by stretched-exponential relaxation due to spatial inhomogeneity, is fit equally well with two exponentials, also of equal amplitudes. We attribute these results to two interstitial muon sites in the unit cell, rather than disorder or other spatial distribution. Further evidence for this attribution is found from agreement between the ratio of the two measured relaxation rates and calculated mean-square local Yb$^{3+}$ dipolar fields at candidate muon sites. Zero-field data can be understood as a combination of two-exponential dynamic relaxation and quasistatic nuclear dipolar fields.
\end{abstract}


\maketitle

\section{INTRODUCTION} \label{sec:intro}

In 1973 Anderson~\cite{Anderson73} proposed an exotic magnetic state in the two-dimensional (2D) triangular-lattice Heisenberg antiferromagnet where local spins are highly entangled with no spin freezing. A clear experimental identification of such a quantum spin liquid (QSL) material is much desired, for its own sake and for potential applications in quantum computation and possible relation to novel phenomena such as high-temperature superconductivity, topological order, and Mott insulators~\cite{Anderson87, Lee08, Balents10, Savary17, Zhou17}. Many materials with geometrically-frustrated lattices have been proposed as candidate spin liquids (see, e.g.,~\cite{Uemura94, Shimizu03, Nakatsuji06, Helton07, Okamoto07, Itou08, Li15SR}). Whether there is a QSL among these candidates remains controversial, however.

YbMgGaO$_4$ is a quasi-2D triangular-lattice antiferromagnet that has attracted considerable attention as a  QSL candidate~\cite{Li15SR, Li15PRL, Li16, Xu16, Shen16, Paddison17, Li17PRL, Li17NC, Shen18, Zhang18prx, Ma18, Li19}. No spin ordering was observed down to 0.06~K from thermodynamic and transport measurements ~\cite{Li15SR, Xu16}. The low-tem\-per\-a\-ture heat capacity exhibits a power-law temperature dependence with an exponent close to the theoretical value of 2/3 for a $U(1)$ QSL\@. An anisotropic spin interaction was observed and discussed as important to stabilize a possible QSL ground state~\cite{Li15PRL}. Inelastic neutron scattering measurements~\cite{Shen16, Paddison17, Li17NC, Li19} provided evidence for a QSL state with a spinon Fermi surface by observing continuous spin excitations. High-field time-domain terahertz spectroscopy measurements~\cite{Zhang18prx} characterized YbMgGaO$_4$ as an easy-plane XXZ antiferromagnet. Theoretical treatments predict a proposed spinon Fermi surface $U(1)$ spin liquid ~\cite{LiYaodong16, LiYaodong17, LiYaodongField17}. 

Whether or not there is a QSL state in YbMgGaO$_4$ is still under debate. Evidence against a QSL includes thermal conductivity measurements that did not observe the expected magnetic excitation contribution down to 0.050~K~\cite{Xu16}. Furthermore, a frequency-dependent cusp was observed at $\sim$0.1~K in ac susceptibility measurements~\cite{Ma18}, suggesting a spin-glass-like transition, although the dc magnetization saturates at low temperatures with no evidence for spin freezing up to 10~kOe~\cite{Li19}. Proposed alternative ground states~\cite{Luo17, Zhu17, Parker18, Kimchi18} involve various forms of ordered or disordered static magnetism, but without definite predictions for transition temperatures. 

Muon spin rotation and relaxation ($\mu$SR)~\cite{Schenck85, *Blundell99, *Brewer03, *Yaouanc11} is a powerful technique to probe local magnetism in solids, and has been widely used to study spin-liquid candidates (e.g.,~\cite{Uemura94, Bert09, Pratt11Science, Clark13, Khuntia16, Li16, Pratt18}). A $\mu$SR study of YbMgGaO$_4$ by Y.~Li \emph{et~al.}~\cite{Li16} showed no static or long-range magnetic order in zero field (ZF) down to 0.048~K, which was taken as evidence for a $U(1)$ QSL ground state. Below 10~K the ZF muon spin polarization~$P_\mu(t)$ exhibited so-called stretched-exponential relaxation~$P_\mu(t) = \exp[-(\lambda^\ast t)^\beta]$, where $\lambda^\ast$ is a characteristic relaxation rate and $\beta < 1$ is the stretching exponent. Stretched exponentials are used as phenomenological models for spatially inhomogeneous distributions of relaxation rates~\cite{Johnston06}. In YbMgGaO$_4$ $\beta \approx 0.95$ at 10~K~\cite{Li16}; the ZF relaxation was nearly exponential. With decreasing temperature $\beta$ decreased considerably, suggesting the onset of spatial inhomogeneity. At 70~mK $\lambda^\ast$ was continuously suppressed by an applied longitudinal field (LF) up to 1.8~kOe. The field dependence of $\beta$ was not reported.

This article reports $\mu$SR experiments on single-crystalline YbMgGaO$_4$ that extend the work of Li \emph{et~al.}~\cite{Li16}. Positive-muon ($\mu^+$) frequency and relaxation measurements were carried out for temperatures down to 22~mK and applied fields up to 10~kOe. Fourier transforms of $\mu$SR data taken in a transverse field (TF)~$H_T = 10$~kOe (Sec.~\ref{sec:TF}) reveal two lines with equal amplitudes but significantly different frequency (Knight) shifts. We attribute the lines to two distinct $\mu^+$ interstitial stopping sites near the two inequivalent oxygen ions in the YbMgGaO$_4$ unit cell, as previously suggested~\cite{Li16}. The corrected Knight shifts~$K_\mathrm{corr}$ are nearly temperature-independent below 1~K, as is the bulk susceptibility~$\chi_\mathrm{mol}$ at this field (Ref.~\cite{Li19}, supplement). The ratio~$K_\mathrm{corr}/\chi_\mathrm{mol}$ is roughly consistent with calculated Yb$^{3+}$ dipolar fields at candidate $\mu^+$ sites.

$\mu^+$ spin relaxation data were taken for longitudinal  fields~$H_L$ up to 8~kOe at 22~mK, 0.20~K, and 5.83~K (Sec.~\ref{sec:LF}), and in ZF over the temperature range~22~mK--5.8~K (Sec.~\ref{sec:ZF}). In light of the TF-$\mu$SR results summarized above, LF asymmetry spectra for $T = 22$~mK and $H_L \geqslant 25$~Oe, where the relaxation is purely dynamic~\cite{Schenck85, *Blundell99, *Brewer03, *Yaouanc11}, were fit by a sum of two exponentials (rates $\lambda_1$ and $\lambda_2$), rather than the stretched exponential used previously~\cite{Li16}. A sum of two exponentials and a stretched exponential are both sub-exponential forms (i.e., exhibit upward curvature on a log-linear plot), and can be difficult to distinguish~\cite{[{See, for example, }] Johnston06}. 

With equal amplitudes of the two exponentials, the observed ratio~$\lambda_2/\lambda_1 = 0.03(1)$ is field-independent and in rough agreement with the ratio of calculated mean-square Yb$^{3+}$ dipolar fields. This is expected if the sub-exponential relaxation is due to differences in coupling fields rather than inhomogeneity in the spin dynamics. It is additional evidence for the two-site scenario. 

For lower $H_L$ the contributions of static nuclear dipolar fields must be considered (stretched-exponential fits alone mix the static and dynamic contributions). Fits of relaxation functions appropriate to this situation~\cite{Schenck85, *Blundell99, *Brewer03, *Yaouanc11} (Sec.~\ref{sec:LF}) to the data then allow determination of $\lambda_1(H_L)$ down to $H_L = 0$. 

The paper is organized as follows. Section~\ref{sec:exp} describes the YbMgGaO$_4$ sample and its crystal structure, briefly summarizes the $\mu$SR technique, and gives details of the experiment. Results of TF, LF, and ZF experiments are reported in Sec.~\ref{sec:results} and are discussed in Sec.~\ref{sec:disc}. Section~\ref{sec:concl} summarizes our conclusions.

\section{EXPERIMENT} \label{sec:exp}

\subsection{Sample} \label{sec:sample}

\begin{figure}[ht] 
 \includegraphics[clip=,width=0.45\textwidth]{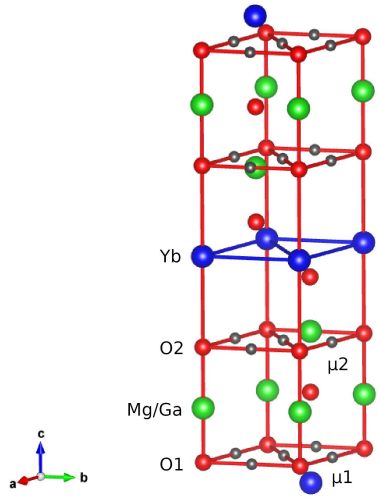}
 \caption{\label{fig:YMGO} Two-thirds of the unit cell of YbMgGaO$_4$~\cite{Li15SR, Li15PRL}. Blue spheres: Yb. Green spheres: Mg/Ga. Red spheres: O. Small gray spheres: candidate $\mu^+$ sites~$\mu$1 and $\mu2$ (see text).}
\end{figure}

Plate-like high-quality single crystals of YbMgGaO$_4$ were grown by a floating-zone method and characterized as reported previously~\cite{Ma18}. Figure~\ref{fig:YMGO} 
shows two-thirds of a hexagonal unit cell of YbMgGaO$_4$ (space group~$R\bar{3}m$, No.~166)~\cite{Li15SR, Li15PRL}. Mg and Ga ions are distributed randomly on the indicated sites, leading to unavoidable local disorder in the structure. There are two inequivalent oxygen locations (O1 and O2) in the unit cell. The triangular 2D Yb$^{3+}$ layers are well separated by two Mg/Ga planes and four O1/O2 planes (Yb$^{3+}$ layer spacing 8.377~\AA), and inter-layer magnetic interaction is weak. 

Implanted positive muons in oxides are likely to stop near O$^{2-}$ ions. To date there are no reported calculations of $\mu^+$ site locations in YbMgGaO$_4$~\cite{[{For a review, see }] Bonfa16}. In lieu of such information, we consider candidate sites that are centered with respect to oxygen nearest neighbors. There are two such crystallographically inequivalent sites ($\mu$1 and $\mu$2), shown in Fig.~\ref{fig:YMGO}. Dipolar magnetic fields from neighboring Yb$^{3+}$ ions and nuclei are unlikely to be very different if the actual $\mu^+$ sites are closer to O$^{2-}$ ions.

\subsection{\boldmath The $\mu$SR technique} 

In a typical $\mu$SR experiment~\cite{Schenck85, *Blundell99, *Brewer03, *Yaouanc11}, $\sim$100\% spin-pol\-ar\-ized positive muons are implanted into the sample and stop at interstitial sites. Each $\mu^+$ precesses in the (in general fluctuating) local field at its site, and at the time of its decay (mean lifetime ${\sim}2.2~\mu$s) emits a positron preferentially in the direction of the $\mu^+$ spin. Positrons are detected by an array of scintillation counters, and a histogram~$N(t)$ of count rate vs time after implantation is obtained from each counter. 

For two identical counters on opposite sides of the sample, the $\mu$SR asymmetry time spectrum~$A(t) = [N_1(t)-N_2(t)]/[N_1(t)+N _2(t)]$ is proportional to the component~$P_\mu(t)$ of the ensemble $\mu^+$ spin polarization~$\mathbf{P}_\mu(t)$ along the axis joining the counters and the sample. Corrections for differences in counter geometries and efficiencies are straightforward~\cite{Schenck85, Blundell99, Brewer03, Yaouanc11}. Unlike its cousin NMR, which, apart from NQR and field-cycling techniques, requires a strong applied magnetic field to polarize the nuclear spins, the polarized $\mu^+$ beam allows $\mu$SR to be carried out in arbitrary fields. 

Experiments may be considered in three classes:
\begin{itemize}
\item In high TF-$\mu$SR, the precession frequency in a paramagnet is shifted from its value \emph{in vacuo}. The fractional frequency shift~$K$ (often called the Knight shift after its discoverer~\cite{Knight49}) and its inhomogeneous distribution (often the main contribution to the linewidth) provide a microscopic characterization of the field-induced static magnetization. 

\item In sufficiently strong LF-$\mu$SR, the resultant of applied and any local static fields is essentially parallel to the initial $\mu^+$ spin orientation. This results in ``decoupling'' of the static fields, and the remaining relaxation is purely dynamic in origin~\cite{Hayano79}. 

\item In ZF- and weak LF-$\mu$SR, the time evolution of $P_\mu(t)$ is due to either or both of two mechanisms: (1)~decay due to dynamic fluctuations of the local fields, and (2)~$\mu^+$ spin precession in the resultant of a distribution of (quasi)static local fields and the applied field, if any (static Kubo-Toyabe (KT) relaxation~\cite{Hayano79}). The latter are normally due to nuclear dipolar fields. 

This situation can be modeled by a dynamically-damped static KT function~$G_\mathrm{KT}(t)$:
\begin{equation} \label{eq:dampedKT}
P_\mu(t) = G_d(t)\, G_\mathrm{KT}(t) \,,
\end{equation}
where the dynamic damping function~$G_d(t)$ is often a simple exponential but can be more complex (as in the present work). Choices of $G_\mathrm{KT}(t)$ are discussed below in Sec.~\ref{sec:allLF}. It is, however, often difficult to disentangle static and dynamic relaxation contributions from ZF-$\mu$SR data alone; this is one reason why strong LF-$\mu$SR is valuable. 

\end{itemize}

\subsection{Experimental details} 

A mosaic of YbMgGaO$_4$ single crystals (total mass $\approx 0.3$~g) was mounted on a pure silver sample holder using diluted GE varnish, with the crystal $c$ axes parallel to the $\mu^+$ beam direction. $\mu$SR experiments were carried out using the dilution refrigerator spectrometer at the M15 beam line of TRIUMF, Vancouver, Canada. Field zeroing to ${\lesssim}10$~mOe was achieved using a variation of the method of Morris and Heffner~\cite{Morris03}. 

Data were taken over the temperature range~22~mK--8~K\@. Due to the large low-tem\-per\-a\-ture specific heat of YbMgGaO$_{4}$~\cite{Xu16}, temperatures were stabilized for 15 minutes before taking data. ZF- and LF-$\mu$SR data were taken with the initial $\mu^+$ ensemble spin polarization parallel to the crystal $c$ axes. For TF-$\mu$SR the $\mu^+$ spins were rotated by ${\sim}90^\circ$ before implantation. Data were analyzed using the Paul Scherrer Institute \textsc{musrfit} fitting program~\cite{Suter12} and the TRIUMF \textsc{physica} programming environment~\footnote{\texttt{http://computing.triumf.ca/legacy/physica/}}.

\section{RESULTS} \label{sec:results}

\subsection{\boldmath Transverse-field $\mu$SR} \label{sec:TF}

\begin{figure}[ht]
 \includegraphics[clip=,width=0.45\textwidth]{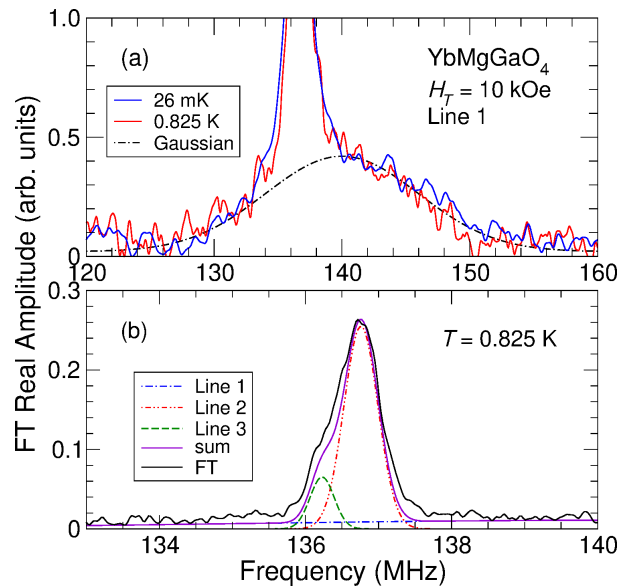}
 \caption{Fourier-transform (FT) $\mu^+$ spectra from YbMgGaO$_4$, $H_T = 10$~kOe. (a)~Wide-frequency FTs at 26~mK (blue) and 0.825~K (red). The narrow peak is distorted (broadened) by strong apodization. (b)~Narrow-frequency FT, $T = 0.825$~K\@. Curves: Fourier transforms of Gaussian terms~1--3 [Eq.~(\ref{eq:KSfit})] and their sum (see text).}
 \label{fig:FT}
\end{figure}

TF-$\mu$SR experiments were carried out in YbMgGaO$_4$ over the temperature range~26~mK--0.825~K in a transverse applied field~$H_T = 10$~kOe. Representative Fourier transforms of TF asymmetry time spectra are shown in Fig.~\ref{fig:FT}.
They exhibit three lines: line 1 [Fig.~\ref{fig:FT}(a)], which is very broad and shifted to higher frequency, and lines~2 and 3 [Fig.~\ref{fig:FT}(b)], which are narrow and overlap considerably. 

The original TF asymmetry time spectra~$A^\mathrm{TF}(t)$ (not shown) were therefore fit by a sum of three Gaussian-damped oscillating components:
\begin{equation} \label{eq:KSfit}
A^\mathrm{TF}(t) = \sum_{i=1}^3 A_i^\mathrm{TF}\exp(-{\textstyle\frac{1}{2}}\sigma_i^2t^2)\cos(\omega_i t+\varphi) \,,
\end{equation}
where a common phase~$\varphi$ was assumed. Parameter values for $T = 26$~mK are listed in Table~\ref{tab:TFparams}. 
\begin{table*} [ht]
\caption{\label{tab:TFparams} Parameters~$A_i^\mathrm{TF}$, $\omega_i$, and $\sigma_i$ from fits of Eq.~(\ref{eq:KSfit}) to $\mu^+$ asymmetry time spectra from YbMgGaO$_4$, $H_T = 10$~kOe, $T = 26$~mK\@. Hyperfine coupling constants~$A_\mathrm{hf}$ and hyperfine fields~$H_\mathrm{hf}$ (experimental and calculated dipolar) are discussed in the text.}
\begin{ruledtabular}
\begin{tabular}{ccccccc}
line~$i$ & $A_i^\mathrm{TF}$ & $\omega_i/2\pi$ & $\sigma_i$ & $A_\mathrm{hf}$ & $H_\mathrm{hf}^{(\mathrm{exp})}$ & $H_\mathrm{hf}^{(\mathrm{dip})}$ \footnote{From lattice sums of nuclear dipolar fields.} \\
 & & (MHz) & ($\mu\text{s}^{-1}$) & (mol/emu) & (kOe/$\mu_B$ per Yb) & (kOe/$\mu_B$ per Yb) \\
\colrule
 1 & 0.140(7) & 140.2(3) & 32.5(1.3) & 0.23(1) & 1.26(6) & $-$1.174 ($\mu$1) \\
 2 & 0.149(5) & 136.760(5) & 1.47(2) & 0.13(1) & 0.73(5) & 0.897 ($\mu$2) \\ 
3 & 0.027(5) & 136.23(1) & 1.04(5) & --& -- & -- \\ 
 \end{tabular}
\end{ruledtabular}
\end{table*}
Amplitudes~$A_1^\mathrm{TF}$ and $A_2^\mathrm{TF}$ from the fits are nearly equal and considerably larger than $A_3^\mathrm{TF}$. The curves in Fig.~\ref{fig:FT} are Fourier transforms of the Gaussian asymmetry terms in Eq.~(\ref{eq:KSfit}), with parameters from the fits: frequencies~$\omega_i/2\pi$, widths~$\sigma_i/2\pi$, and areas proportional to $A_i$. Their sum is normalized to the experimental Fourier transform in Fig.~\ref{fig:FT}(b). 

One of the three lines arises from muons that miss the sample and stop in the silver sample holder, and the other two are from the sample. Line~3, which is the narrowest, is assumed to be the ``silver'' signal, and $\omega_3/2\pi$ is used as the reference frequency for the Knight shifts of the other two components. Additional evidence for this attribution is discussed below in Sec.~\ref{sec:hiLF}. The sample lines are numbered in order of their shifts (Fig.~\ref{fig:FT}), consistent with the $\mu^+$ site numbering in order of their distance from the nearest Yb$^{3+}$ layer (Fig.~\ref{fig:YMGO}).

In order for Knight shifts to reflect local magnetism, a correction must be made for the contributions to the raw shift~$K_\mathrm{raw}$ from macroscopic Lorentz and demagnetizing fields~~\cite{Carter77, *Akashin92}:
\begin{equation} \label{eq:Ldcorr}
K_\mathrm{corr} = K_\mathrm{raw} - A_\mathrm{Ld}\,\chi_\mathrm{mol} \,,
\end{equation}
where $A_\mathrm{Ld} = 4\pi\left(\textstyle{\frac{1}{3}} - D\right)/v_\mathrm{mol}$, $\chi_\mathrm{mol}$ is the molar bulk susceptibility, $D$ is the sample demagnetization coefficient, and $v_\mathrm{mol}$ is the molar volume (50.6~cm$^3$/mol for YbMgGaO$_4$). For field normal to the thin mosaic samples, $D $ is estimated to be 0.8(1), so that $A_\mathrm{Ld} = -0.12(2)$~mol~emu$^{-1}$. This correction can be important in $\mu$SR shift measurements. 

The rates~$\sigma_i$ in Table~\ref{tab:TFparams} are faster than the dynamic rates discussed below in Secs.~\ref{sec:LF} and \ref{sec:ZF}, particularly for line 3. $\sigma_i$ is therefore attributed to static distributions of TF frequencies, most likely related to the crystalline-electric-field randomness observed in neutron scattering~\cite{Li17PRL}, which lead to rms widths~$\delta K = \sigma/\omega_3$ of Knight shift distributions. We do not fully understand the large difference between $\sigma_1$ and $\sigma_2$.

\begin{figure}[ht]
 \includegraphics[clip=,width=0.45\textwidth]{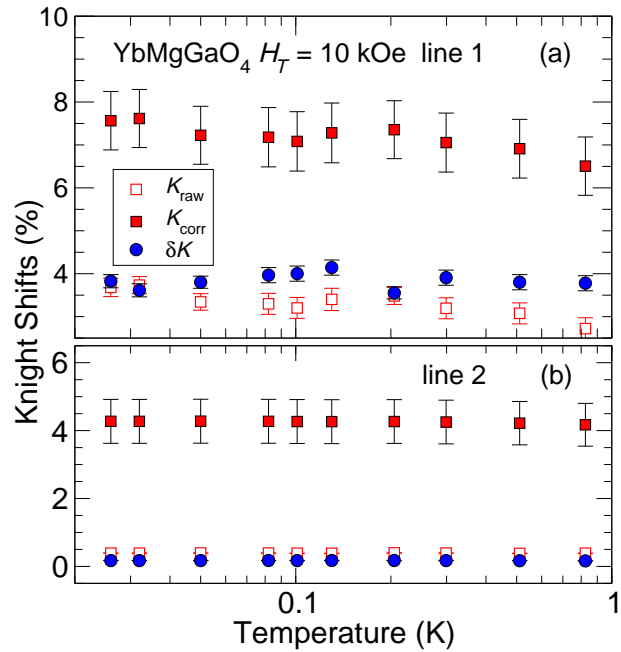}
 \caption{\label{fig:KSvsT} Temperature dependence of $\mu^+$ Knight shifts in YbMgGaO$_4$ (Fig.~\ref{fig:FT}). (a)~Line 1. (b)~Line~2. Open squares: raw shifts~$K_\mathrm{raw}$. Filled squares: corrected shifts~$K_\mathrm{corr}$ (see text). Circles: rms distribution widths~$\delta K = \sigma/\omega_3$ (Table~\ref{tab:TFparams}).}
\end{figure}

The temperature dependencies of $K_\mathrm{raw}$, $K_\mathrm{corr}$, and $\delta K$ for lines 1 and 2 are given in Fig.~\ref{fig:KSvsT}.
The error in $K_\mathrm{corr}$ is largely due to uncertainty in $A_\mathrm{Ld}$ [Eq.~(\ref{eq:Ldcorr})]. Over the temperature range of the data both the Knight shifts and $\chi_\mathrm{mol}(T) = M(T)/H$ in 10~kOe~\cite{Li19} are practically constant. 

The relation between the Knight shift and $\chi_\mathrm{mol}$ is often analyzed using the Clogston-Jaccarino relation~\cite{Clogston61}
\begin{equation} \label{eq:clogjacc}
K_\mathrm{corr}(T) = K_0 + A_\mathrm{hf}\,\chi_\mathrm{mol}(T) \,,
\end{equation}
where $A_\mathrm{hf}$ is the hyperfine coupling constant and temperature is an implicit variable. In the present case there is not enough temperature dependence to either quantity for this procedure to be reliable. Instead, we assume $K_0$ in Eq.~(\ref{eq:clogjacc}) is negligible, as would be expected in a magnetic insulator, so that $A_\mathrm{hf} = K_\mathrm{corr}/\chi_\mathrm{mol}$. The hyperfine field $H_\mathrm{hf}$ (the local field at a $\mu^+$ site per $\mu_B$ of paramagnetic moment on the local-moment sites) is given by $H_\mathrm{hf} = N_A\mu_BA_\mathrm{hf}$, where $N_A$ is Avogadro's number. 

Values of $A_\mathrm{hf}$ and $H_\mathrm{hf}$ for the two lines are given in Table~\ref{tab:TFparams}. Also included in Table~\ref{tab:TFparams} are calculated Yb$^{3+}$ dipolar hyperfine fields at each of the two candidate $\mu^+$ sites discussed in Sec.~\ref{sec:sample} (Appendix~\ref{sec:diplattsums}); local-moment/$\mu^+$ coupling fields are expected to be predominantly dipolar in insulators~\cite{Schenck85}. The calculated and measured values are of the same order of magnitude, but with a discrepancy in sign for line~1. This is not understood, but it may mean that the location of site $\mu$1 is incorrect.
 
\subsection{\boldmath Longitudinal-field $\mu$SR} \label{sec:LF}

\subsubsection{\label{sec:hiLF} $H_L \geqslant 25$~Oe; two-exponential relaxation} \label{sec:2exp}

We first consider data for $H_L$ strong enough to decouple any nuclear dipolar fields at $\mu^+$ sites. Then the relaxation is purely dynamic, and its analysis is simplified.

The observation of two sample lines in TF-$\mu$SR and their identification with distinct $\mu^+$ sites motivates fitting the data to a sum of two exponentials 
\begin{equation} \label{eq:2exp}
A_d(t) = A_0 \sum_{i=1}^2 f_i\exp(-\lambda_it) + A_\mathrm{Ag} \,,
\end{equation}
rather than the stretched exponential used previously~\cite{Li16}. Here $A_0$ is the initial sample asymmetry, $\lambda_1$ and $\lambda_2$ are the fast and slow relaxation rates, respectively, $A_\mathrm{Ag}$ is the asymmetry of the background silver signal noted above in Sec.~\ref{sec:TF}, and $f_1 + f_2 = 1$. Good fits to the data (Appendix~\ref{sec:2expfits}) were obtained using Eq.~(\ref{eq:2exp}) with $f_2$ fixed at 0.52 from the amplitudes of the TF lines (Table~\ref{tab:TFparams}). This is \emph{a posteriori} justification for the assumption that lines~1 and 2 arise from the sample; the amplitude of line~3 is too small to produce significantly sub-exponential relaxation.

\begin{figure}[ht]
 \includegraphics[clip=,width=0.45\textwidth]{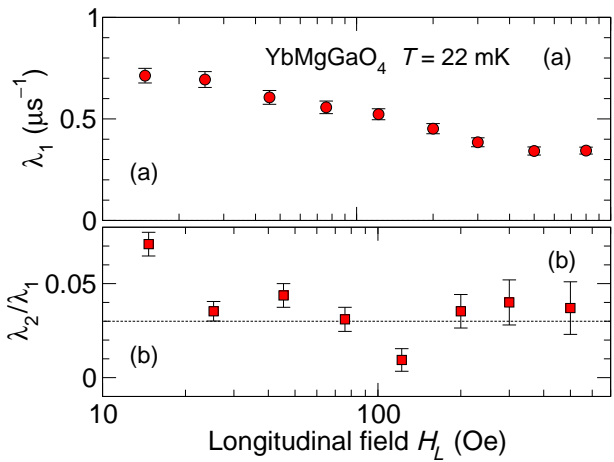}
 \caption{\label{fig:2exp-LF} Field dependencies of dynamical relaxation rates from two-exponential fits to LF-$\mu$SR data in YbMgGaO$_4$, $T = 22$~mK\@. (a)~Rapid relaxation rate~$\lambda_1$. (b)~Ratio of rates~$\lambda_2/\lambda_1$.}
 \end{figure}

The field dependencies of the fast rate~$\lambda_1$ and the ratio~$\lambda_2/\lambda_1$ at $T = 22$~mK are shown in Fig.~\ref{fig:2exp-LF} for intermediate fields.
For $25~\text{Oe} \lesssim H_L \lesssim 500~\text{Oe}$, the ratio~$\lambda_2/\lambda_1 = 0.03(1)$ is basically constant, and agrees roughly with the value~0.021 from a lattice sum of uncorrelated mean-square Yb$^{3+}$ dipolar fields at the candidate $\mu^+$ sites (Appendix~\ref{sec:diplattsums}). This is expected in the fast-fluctuation (motionally narrowed) limit if the fluctuation rates at the two sites are the same and the correlation length for the fluctuations is short~\cite{Miao16}.

Its small value makes $\lambda_2$ difficult to measure accurately. Furthermore, for $H_L \gtrsim 800$~Oe, $\lambda_1$ becomes too slow to permit distinguishing the two components from data in the $\mu$SR time window (Appendix~\ref{sec:2expfits}), so that $\lambda_2$ cannot be determined at high fields. This together with its observed proportionality to $\lambda_1$ (Fig.~\ref{fig:2exp-LF}) justifies consideration only of the latter in the following.

The increase of $\lambda_2/\lambda_1$ at 14~Oe (Fig.~\ref{fig:2exp-LF}) signals the onset of static relaxation at low fields. As discussed below in Sec.~\ref{sec:disc}, in zero and low fields a static (Gaussian) local-field component reduces the sub-exponential character of the relaxation, thereby increasing $\lambda_2/\lambda_1$ in the now-incorrect two-exponential fit. 

\subsubsection{\label{sec:allLF} $0 \leqslant H_L \lesssim 8$~kOe; two-exponential + KT static relaxation} 

In the presence of static nuclear dipolar local fields, both static and dynamic processes contribute to $\mu^+$ relaxation at low LF\@. For the two-site scenario, the asymmetry function becomes
\begin{eqnarray} \label{eq:2expKT}
A(H_L,t) & = & A_0\sum_{i=1}^2 f_i \exp(-\lambda_i t) G_\mathrm{KT}(H_L,\Delta_i,t) \nonumber \\
& & + A_\mathrm{Ag} \,,
\end{eqnarray}
where $G_\mathrm{KT}(H_L,\Delta,t)$ is the static KT function for Gaussian local-field distributions~\cite{Hayano79}. Here $\Delta/\gamma_\mu$ is the rms width of the Gaussian field distribution, where $\gamma_\mu/2\pi = 1.3553 \times 10^4~\text{Hz~G}^{-1}$ is the $\mu^+$ gyromagnetic ratio. In YbMgGaO$_4$ we attribute this distribution to nuclear dipolar fields: $\Delta_i = \Delta_i^\mathrm{nuc}$.

Equation~(\ref{eq:2expKT}) is also valid for $H_L = 0$, where it is most sensitive to the values of the $\Delta_i^\mathrm{nuc}$. These were determined by fits of Eq.~(\ref{eq:2expKT}) to ZF data, discussed in more detail in Sec.~\ref{sec:ZF}. For fits to the LF data, $\lambda_2/\lambda_1$ and the $\Delta_i^\mathrm{nuc}$ were fixed at 0.03 and their ZF values, respectively. 

\begin{figure}[ht]
\includegraphics[clip=,width=0.45\textwidth]{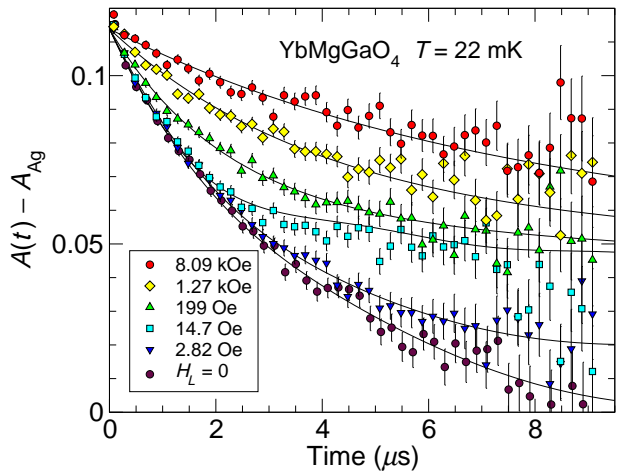}
 \caption{\label{fig:22mKpol} LF-$\mu$SR asymmetry spectra in YbMgGaO$_4$, $T = 22$~mK, $0 \leqslant H_L \lesssim 8$~kOe, after subtraction of a non-relaxing signal from the sample holder. Curves: fits of Eq.~(\ref{eq:2expKT}) to the data.}
\end{figure}

Figure~\ref{fig:22mKpol}
shows asymmetry spectra for representative values of $H_L$. At low fields the field dependence is dominated by decoupling of the static relaxation, with its characteristic field independence at early times. At higher fields the predominantly dynamic relaxation becomes slower with increasing field as previously reported~\cite{Li16}.

\begin{figure}[ht]
 \includegraphics[clip=,width=0.45\textwidth]{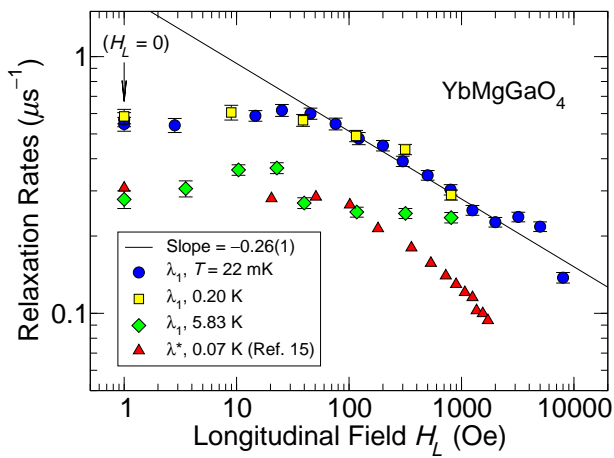}
 \caption{\label{fig:LFrate} Field dependencies of dynamic relaxation rates from fits to LF-$\mu$SR data in YbMgGaO$_4$, $T = 22$~mK (circles), 0.20~K (squares), and 5.83~K (diamonds). $\lambda_1$: fast rate from two-exponential-damped static Gaussian KT fits. $\lambda^\ast$: stretched-exponential rate from Ref.~\cite{Li16}, $T = 70$~mK (triangles). Straight line: power-law fit to 22-mK data for $25~\text{Oe} \leqslant H_L \leqslant 2$~kOe.}
\end{figure}

The field dependencies of $\lambda_1$ at temperatures of 22~mK, 0.20~K, and 5.83~K are shown in Fig.~\ref{fig:LFrate}, 
together with stretched-exponential rates~$\lambda^\ast$ for $T = 70$~mK from Ref.~\cite{Li16}. The $\lambda_1$ values for 22~mK are in agreement with the two-exponential fit results for $H_L \gtrsim 45$~Oe, since $G_\mathrm{KT}(H_L,\Delta,t) \to 1$ for $H_L \gg \Delta/\gamma_\mu$. $\lambda_1(H_L)$ and $\lambda^\ast(H_L)$ follow the same trend, but do not agree quantitatively because the assumed relaxation functions are different. 

The straight line in Fig.~\ref{fig:LFrate} is a power-law fit to the 22-mK data for $25~\text{Oe} \leqslant H_L \leqslant 2$~kOe, almost two decades of field values. Power-law behavior is a form of time-field scaling~\cite{Keren00}, since $P_\mu(H_L,t) = P_\mu[\lambda_i(H_L) t] = P_\mu(H_L^{-x}t)$ for $\lambda_i \propto H_L^{-x}$. Time-field scaling is in general indicative of long-lived spin correlations with a power-law divergence in the noise spectrum as $\omega \to 0$. This result is consistent with the conclusion of Li \emph{et~al.}~\cite{Li16} from a more detailed analysis.

At 5.83~K the rates are reduced, and exhibit a plateau for $H_L \gtrsim 40$~Oe. The rather abrupt crossover may be to a normal paramagnetic state, in which local-moment fluctuations are independent of both field and temperature~\cite{Moriya56}.

\subsection{\boldmath Zero-field $\mu$SR} \label{sec:ZF}

\subsubsection{ZF asymmetry spectra}

\begin{figure}[ht]
 \includegraphics[clip=,width=0.45\textwidth]{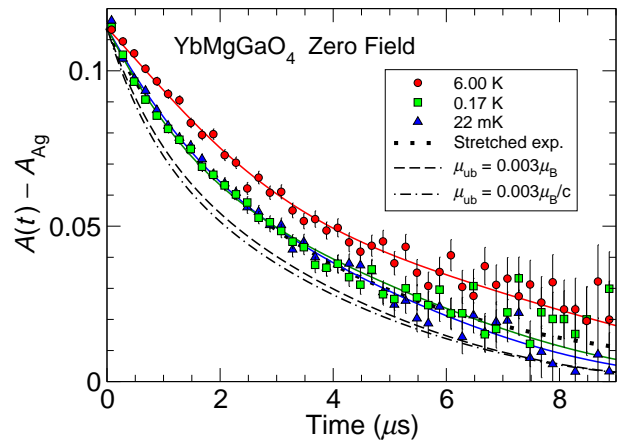}
 \caption{\label{fig:ZF-pol} ZF asymmetry spectra from YbMgGaO$_4$, $T = 22$~mK (triangles), 0.17~K (squares) and 6.00~K (circles), after subtraction of a non-relaxing signal from the silver sample holder. Solid curves: fits of Eq.~(\ref{eq:2expKT}) to the data. Dotted curve: stretched-exponential fit for 22~mK\@. Dashed curve: fit spectrum for 22~mK with additional static Gaussian relaxation, moments~$0.003\,\mu_B$ per Yb ion. Dash-dot curve: fit spectrum for 22~mK with additional static spin-glass relaxation, moment-concentration product~$\mu = 0.003\,\mu_B/c$ per Yb ion (see text).}
\end{figure}

Extraction of parameters in Eq.~(\ref{eq:2expKT}) from fits in ZF and weak LF that include static nuclear dipolar relaxation is difficult, because the data are smooth and without distinguishing features (Figs.~\ref{fig:22mKpol} and \ref{fig:ZF-pol}). Fits with all parameters free lead to excessive scatter, so that it is necessary to fix as many parameters as possible. 

Fixing $f_2 = 0.52$ and $\lambda_2/\lambda_1 = 0.03$ have been discussed above. The $\Delta_i^\mathrm{nuc}$ were then allowed to vary, leading to fits (not shown) with excessive scatter but reasonably temperature-independent $\Delta_i^\mathrm{nuc}$ values below $\sim$0.5~K\@. The data were finally refit with $\Delta_i^\mathrm{nuc}$ fixed at their averages~$\Delta_1^\mathrm{nuc} = 0.21(2)~\mu\text{s}^{-1}$ and $\Delta_2^\mathrm{nuc} = 0.15(1)~\mu\text{s}^{-1}$. No temperature dependence of the nuclear dipolar relaxation is expected in the absence of $\mu^+$ diffusion, which is unlikely at low temperatures~\cite{Schenck85}. 

ZF-$\mu$SR asymmetry spectra for $T = 22$~mK, 0.17~K, and 6.00~K are shown in Fig.~\ref{fig:ZF-pol}, 
again with the silver background signal subtracted. The solid curves are fits of Eq.~(\ref{eq:2expKT}) to the data. The temperature dependence of the overall relaxation over this range is weak because of the temperature-independent static KT contribution, and there is essentially no temperature dependence between 0.17~K and 22~mK\@. We observe no oscillating signal due to long-range magnetic order and no initial rapid relaxation or asymmetry loss due to short-range magnetic order~\cite{Miao16} down to 22~mK\@. The dotted curve is a stretched-exponential fit for $T = 22$~mK, illustrating the similarity between it and the two-exponential fit. The dashed and dash-dot curves are discussed in the next section.

\subsubsection{Limits on static magnetism} \label{sec:maglimits}

To determine upper bounds on static magnetism in YbMgGaO$_4$, two scenarios are considered: equal (small) moments on all Yb$^{3+}$ ions (moment concentration~$c = 1$), and the dilute spin-glass limit for small $c$. For each of these we begin with the ZF data and fit to Eq.~(\ref{eq:2expKT}) for $T = 22$~mK\@. Upper bounds~$\mu_\mathrm{ub}$ on the static moment per Yb ion are chosen, and the corresponding calculated asymmetry spectra are compared with the data. The upper bounds are then adjusted to yield well-resolved increases in relaxation. The results are shown in Fig.~\ref{fig:ZF-pol} for the $c = 1$ and dilute limits by the dashed and dash-dot curves, respectively.

For $c = 1$ the Yb$^{3+}$ dipolar field contribution from disordered Yb$^{3+}$ moments is Gaussian~\cite{Uemura85}, and additional ZF relaxation below a transition temperature increases the values of the Gaussian rates $\Delta_i$ in Eq.~(\ref{eq:2expKT})~\footnote{Long-range magnetic order gives rise to oscillation, which for low frequency varies quadratically with time and cannot be distinguished from a slow Gaussian. The disordered estimate therefore also applies to this case.}. Gaussian rms field widths~$(\Delta_i^\mathrm{(0)}/\gamma_\mu)/\mu_B$ per Yb ion were calculated from dipole-field lattice sums (Appendix~\ref{sec:diplattsums}) at the candidate $\mu^+$ sites~$\mu$1 and $\mu$2 of Fig.~\ref{fig:YMGO}, yielding $\Delta_1^\mathrm{(0)} = 147~\mu\text{s}^{-1}/\mu_B$ and $\Delta_2^\mathrm{(0)} = 46~\mu\text{s}^{-1}/\mu_B$. 

An upper bound~$\mu_\mathrm{ub}$ on the static moment per Yb ion is found by comparing Eq.~(\ref{eq:2expKT}) with increased rates~$\Delta_i' = \left[(\Delta_i^\mathrm{nuc})^2 + (\mu_\mathrm{ub}\Delta_i^\mathrm{(0)})^2\right]^{1/2}$ to the data. The choice~$\mu_\mathrm{ub} = 0.003\,\mu_B$ per Yb ion results in the dashed curve in Fig.~\ref{fig:ZF-pol}, which exhibits a well-resolved increase. For comparison, dc susceptibility measurements in YbMgGa)$_4$~\cite{Ma18, Li19}, which, like $\mu$SR, probe the bulk magnetization, yield paramagnetic Curie-Weiss and saturated moments $\sim 3\,\mu_B$ per Yb ion.

We next consider dilute (disordered) static Yb$ ^{3+}$ moments. The Lorentzian ZF KT relaxation function~\cite{Uemura85}
\begin{equation} \label{eq:Lor}
G_\mathrm{KT}^\mathrm{Lor}(a.t) = \frac{2}{3} + \frac{1}{3}(1 - at)\exp(-at)
\end{equation}
is expected from dipolar coupling to a static random spin system in the dilute limit. Here $a/\gamma_\mu$ is the half-width of the Lorentzian dipolar field distribution~\cite{Uemura85};
\begin{equation} \label{eq:LKT}
a = (\pi/2)^{1/2}c \mu \Delta^\mathrm{(0)}\,, 
\end{equation}
where $\mu$ is the moment per (dilute) Yb ion per Bohr magneton. The resulting relaxation function is derived from Eq.~(\ref{eq:2expKT}) by multiplying each term by $G_\mathrm{KT}^\mathrm{Lor}(a_i,t)$. We choose a value~$\mu_\mathrm{ub} = 0.003/c$ (e.g., for concentration~$c = 0.01$ $\mu_\mathrm{ub} = 0.3\mu_B$ per Yb ion) and determine the $a_i$ from Eq.~(\ref{eq:LKT}), resulting in the dash-dot curve in Fig.~\ref{fig:ZF-pol}. Again, the increase is well resolved. 

Somewhat surprisingly, the sample-average $c = 1$ and dilute bounds are identical. The results cannot be taken as quantitative, however, since they involve the assumed $\mu^+$ sites and do not employ a well-defined resolution criterion. They are solely intended to demonstrate the basic sensitivity of the $\mu$SR technique to weak static magnetism.

\subsubsection{Temperature dependence of zero-field rates}

\begin{figure}[ht]
 \includegraphics[clip=,width=0.45\textwidth]{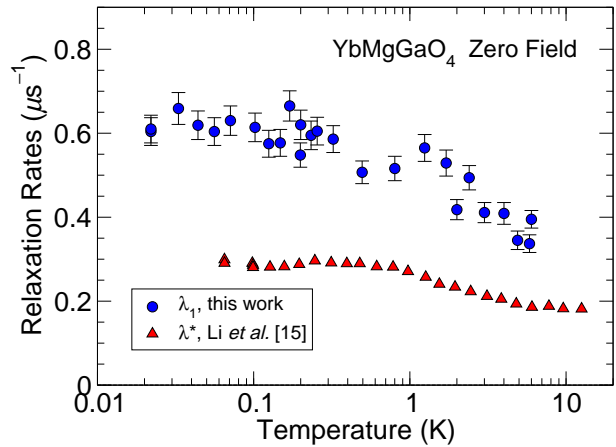}
 \caption{\label{fig:ZFrate} Temperature dependencies of ZF $\mu^+$ relaxation rates in YbMgGaO$_4$ from fits of Eq.~(\ref{eq:2expKT}) to the data. Circles: this work, from fits of Eq.~(\ref{eq:2expKT}) to the data. Triangles: results of Li \emph{et al.}~\cite{Li16}.}
\end{figure}

Figure~\ref{fig:ZFrate} 
shows the temperature dependence of the ZF dynamic relaxation rate~$\lambda_1(T)$ (circles) from fits of Eq.~(\ref{eq:2expKT}) to the data discussed above, and also gives $\lambda^\ast(T)$ from Li \emph{et al.}~\cite{Li16} (triangles). With decreasing temperature, the rates increase and saturate below $\sim$0.5~K, exhibiting persistent spin dynamics down to the lowest temperatures. The scatter in $\lambda_1(T)$ is mainly due to the fact that in ZF the dynamic relaxation is only a fraction of the total relaxation. 

The difference between $\lambda_1(T)$ and $\lambda^\ast(T)$ is again a consequence of the different assumed relaxation functions. Furthermore, as noted above, in ZF and weak LF ``pure'' stretched-exponential fits [i.e., without the factor~$G_\mathrm{KT}(H_L,\Delta,t)$ of Eq.~(\ref{eq:2expKT})] mix the static and dynamic contributions.

\section{DISCUSSION} \label{sec:disc}

Our $\mu$SR study extends previous results~\cite{Li16} concerning the nature of the ground state and low-lying excitations in YbMgGaO$_4$. The observation of the same field dependence~$\lambda_1(H_L)$ at 22~mK and 0.20~K for fields up to 0.8~kOe (Fig.~\ref{fig:LFrate}) complements the previously-reported temperature independence of ZF $\mu^+$ relaxation below $\sim$0.4~K~\cite{Li16}, and is strong evidence that the system is in a temperature-independent state over this region in the $H$-$T$ phase diagram. 

\subsection{Limits on spin freezing}

$\mu$SR experiments are extremely sensitive to static magnetism, ordered or disordered~\cite{Schenck85}, in the present case with an upper limit~${\sim}10^{-3}\mu_B$ per Yb ion on any sample-average static Yb$^{3+}$ moment down to 22~mK (Sec.~\ref{sec:maglimits}). This is a strong constraint on proposals involving magnetic ground states~\cite{Luo17, Zhu17, Parker18, Kimchi18}, although in the disordered local-singlet valence solid~\cite{Kimchi18} the ``ultimate fate of the \dots defect spins is unknown.''

The lack of spin freezing is in apparent contradiction with the observation of a cusp in the low-field ac susceptibility of YbMgGaO$_4$ at $\sim$0.1~K~\cite{Ma18}, and the conclusion that a spin-glass-like transition occurs there. We note that suppression of static order by the muon itself, a possible explanation of this discrepancy, has been shown to be unlikely~\cite{Keren04}. We also note that large static Yb$^{3+}$ moments in a macroscopic spurious phase would result in a fraction of the ZF asymmetry spectrum with oscillations or rapid relaxation. This is not seen at the level of a few percent. The resolution of these issues is not clear, and more work will be required to elucidate the impact of the present results on the QSL debate.

\subsection{Characterization of sample inhomogeneity, consequences for ZF results}

Our $\mu$SR results suggest reinterpretation of some of the previous work. The observation of two equal-amplitude lines in the TF-$\mu$SR data is evidence for the two $\mu^+$ stopping sites in the two inequivalent O$^{2-}$ lattice planes. This, rather than, for example, Mg$^{2+}$/Ga$^{3+}$ site mixing~\cite{Li17PRL}, is the main source of inhomogeneity in the dynamic relaxation, and strongly suggests a two-exponential analysis of LF- and ZF-$\mu$SR data [Eqs.~(\ref{eq:2exp}) and (\ref{eq:2expKT})] rather than a stretched-exponential analysis.

The fact that good fits can be obtained with either two-exponential or stretched-expo\-nen\-tial functions is not surprising. In general, the relaxation function~$P(t)$ associated with a spatially inhomogeneous distribution~$P(\lambda)$ of rates~$\lambda$ is given by the Laplace transform
\begin{equation} \label{eq:Laplace}
P(t) = \int_{-\infty}^\infty P(\lambda)\exp(-\lambda t)\,d\lambda \,.
\end{equation}
Two-exponential and stretched-exponential functions are effectively two choices of $P(\lambda)$. 

Solving Eq.~(\ref{eq:Laplace}) for $P(\lambda)$ is a mathematically ill-posed problem~\cite{Press92}; $P(t)$ is not sensitive to the form of $P(\lambda)$. The difficulty is greater when the range of the independent variable is limited, as it is for $\mu$SR asymmetry spectra by the muon lifetime~\footnote{This is the problem in the present case. At late enough times, the sum of two exponentials with very different rates differs significantly from a stretched exponential}. We also note that for $P(t) = \exp[-(\lambda^\ast t)^\beta]$, the relation of $\lambda^\ast$ to $P(\lambda)$ depends on $\beta$; $\lambda^\ast$ is not simply an average rate, and one should not compare values of $\lambda^\ast$ for different values of $\beta$~\cite{Johnston06}.

The temperature dependence of $\beta$ in fits to ZF data is also consistent with the two-exponential scenario. (Quasi)static relaxation in ZF and weak LF by nuclear dipolar fields should not be ignored, although the lack of structure in the $\mu^+$ asymmetry spectra and the limitation imposed by the muon lifetime make the analysis difficult. We argue that the reported increase of $\beta$ with increasing temperature in ZF stretched-exponential fits is not a temperature-dependent effect of the spin dynamics~\cite{Li16}. Instead, the decrease of ZF dynamic relaxation with increasing temperature (Fig.~\ref{fig:ZFrate}) increases the importance of the temperature-independent ``super-exponential'' (Gaussian) nuclear dipolar field distribution. This renders the overall ZF relaxation function less sub-exponential at high temperatures, leading to $\beta \approx 1$ at $\sim$10~K~\cite{Li16}.

\section{CONCLUSIONS} \label{sec:concl}

We have carried out $\mu$SR experiments on single-crystalline samples of the candidate spin liquid YbMgGaO$_4$, which complement and extend earlier work~\cite{Li16}. The spatial inhomogeneity that gives rise to sub-exponential $\mu^+$ spin relaxation is attributed to two distinct $\mu^+$ stopping sites in the unit cell, rather than Mg$^{2+}$/Ga$^{3+}$ site mixing or other local disorder. With an upper bound of ${\sim}3 \times 10^{-3}\mu_B$/Yb ion, no static magnetism, ordered or disordered, is observed down to 22~mK, a factor of two lower in temperature than previous measurements. The field dependence of the $\mu^+$ spin relaxation is the same in functional form and relaxation rate at 22~mK and 0.20~K up to at least 1~kOe. These results indicate persistent (i.e.,temperature-independent) spin dynamics with no spin freezing for this field and temperature range, thus extending the previous ZF results~\cite{Li16} over a substantial region of the $H$-$T$ phase diagram.

\begin{acknowledgments}
We wish to thank Z. Ma for providing single crystals of YbMgGaO$_4$, and S. Y. Li for fruitful discussions. We are very grateful to the TRIUMF CMMS $\mu$SR support staff for their help during the experiments. This research was supported by the National Key Research and Development Program of China (Nos.~2016YFA0300503 and 2017YFA0303104), the National Natural Science Foundation of China under Grant No.~11774061, and Shanghai Municipal Science and Technology Major Project (Grant No. 2019SHZDZX01). Research at the University of California, Riverside (UCR) was supported by the UCR Academic Senate.
\end{acknowledgments}

\appendix

\section{Dipolar lattice sums} \label{sec:diplattsums}

The dipolar lattice-sum calculation is straightforward. The software sets up a general orthorhombic lattice (with a non-primitive unit cell for a triangular structure such as that of YbMgGaO$_4$). 1-$\mu_B$ local moments are placed at lattice sites in the magnetic structure (with or without a net moment in the unit cell) and candidate $\mu^+$ stopping sites in the unit cell are designated. For each $\mu^+$ site, dipolar fields from the local moments are summed between an inner cutoff radius (if desired) and an outer Lorentz sphere radius~$R_\mathrm{Lor}$. The value of $R_\mathrm{Lor}$ is chosen so that larger values have negligible effect; typically $R_\mathrm{Lor} = 200$~\AA, containing $10^5$--$10^6$ moments, is sufficient. For a paramagnet the moments are all in the same direction, induced by the applied field, and the shift is given by the resultant field in that direction. Mean-square fields, which are involved in fluctuation-induced dynamic relaxation, are calculated by summing the squared magnitudes of the individual moment fields. The calculated fields are in units of [field] per $\mu_B$ local moment. No Lorentz or demagnetizing corrections are made.
 
\section{Two-exponential fits} \label{sec:2expfits}

\begin{figure}
 \includegraphics[clip=,width=0.45\textwidth]{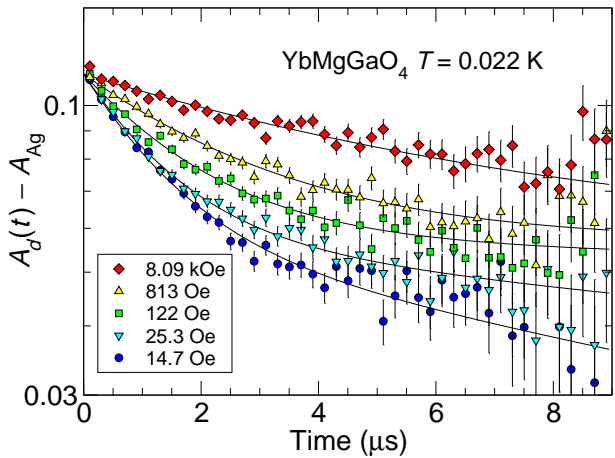}
 \caption{\label{fig:2asy-pol} Log-linear plots of LF asymmetry spectra from YbMgGaO$_4$, $T = 22$~mK, after subtraction of a non-relaxing signal from the silver sample holder. Curves: fits of Eq.~(\ref{eq:2exp}) to the data. The visibly faster slow-component relaxation at 14.7~Oe is attributed to static nuclear dipolar fields.}
\end{figure}

Figure~\ref{fig:2asy-pol} 
shows log-linear plots of representative LF asymmetry spectra at 22~mK, $14.7~\text{Oe} \leqslant H_L \leqslant 8.09$~kOe. Upward curvature is clearly visible. The curves are two-exponential fits to Eq.~(\ref{eq:2exp}) with the slowly-relaxing fraction~$f_2$ fixed at 0.52 (see main text). The fits are good at all fields, but for $H_L = 14.7$~Oe the upward curvature is reduced. As discussed in Secs.~\ref{sec:2exp} and \ref{sec:disc}, this is expected at low enough LF to permit static relaxation by nuclear dipolar fields. For $H_L \gtrsim 1$~kOe, the overall relaxation has slowed enough so that the late-time slow exponential is no longer visible in the experimental time window.


%

\end{document}